\documentclass[runningheads]{llncs}
\usepackage{graphicx}
\usepackage{amsmath}

\usepackage[labelfont=bf]{caption}                         
\usepackage[subrefformat=parens]{subfig}
\usepackage{enumitem}  
\usepackage{amssymb}
\usepackage[misc]{ifsym}
\usepackage{graphicx}
\usepackage{threeparttable}
\usepackage{ntheorem}
\usepackage{color}
\usepackage{colortbl}
\usepackage{multirow}
\usepackage{amsmath}
\usepackage{comment}
\usepackage{bm}                                            
\usepackage{etoolbox}                                      
\usepackage{url}
\usepackage{nth}
\usepackage{cite}
\usepackage{balance}
\usepackage[bookmarks=false]{hyperref}                     %
\usepackage{courier}                                       
\usepackage{listings}                                      
\usepackage[]{algpseudocode}                               
\algrenewcommand\textproc{\texttt}
\makeatletter\let\c@float@type\relax\makeatother
\makeatletter\let\float@addtolists\relax\makeatother
\usepackage{algorithm}
\hypersetup{
    colorlinks = true,
    citecolor  = blue,
    linkcolor  = blue,
	urlcolor   = blue,
}

\newcommand{\tabincell}[2]{                                
	\begin{tabular}{@{}#1@{}}#2\end{tabular}
}

\makeatletter
\newcommand{\thickhline}{%
	\noalign {\ifnum 0=`}\fi \hrule height 1pt
	\futurelet \reserved@a \@xhline
}

\graphicspath{{./}}

\begin{document}

\title{RSANet: Recurrent Slice-wise Attention Network for Multiple Sclerosis Lesion Segmentation}

\author{Hang Zhang\inst{1,3} \Letter \and
Jinwei Zhang \inst{2,3} \and
Qihao Zhang \inst{2,3} \and
Jeremy Kim \inst{4} \and
Shun Zhang \inst{3} \and
Susan A. Gauthier \inst{3} \and
Pascal Spincemaille \inst{3} \and
Thanh D. Nguyen \inst{3} \and
Mert Sabuncu \inst{1,2} \and
Yi Wang \inst{2,3}
}

\authorrunning{H. Zhang et al.}

\institute{School of Electrical and Computer Engineering, Cornell Univeristy \email{hz459@cornell.edu} \and
Meinig School of Biomedical Engineering, Cornell University \and
Department of Radiology, Weill Medical College of Cornell University \and
Hunter College Campus Schools
}

\maketitle

\begin{abstract}

Brain lesion volume measured on T2 weighted MRI images is a clinically important disease marker in multiple sclerosis (MS). 
Manual delineation of MS lesions is a time-consuming and highly operator-dependent task, which is influenced by lesion size, shape and conspicuity. 
Recently, automated lesion segmentation algorithms based on deep neural networks have been developed with promising results. 
In this paper, we propose a novel recurrent slice-wise attention network (RSANet), which models 3D MRI images as sequences of slices and captures long-range dependencies through a recurrent manner to utilize contextual information of MS lesions.
Experiments on a dataset with 43 patients show that the proposed method outperforms the state-of-the-art approaches.
Our implementation is available online at \url{https://github.com/tinymilky/RSANet}


\end{abstract}

\section{Introduction}

\begin{figure}[!t]
	\centering
	\subfloat{\includegraphics[width=.22\columnwidth]{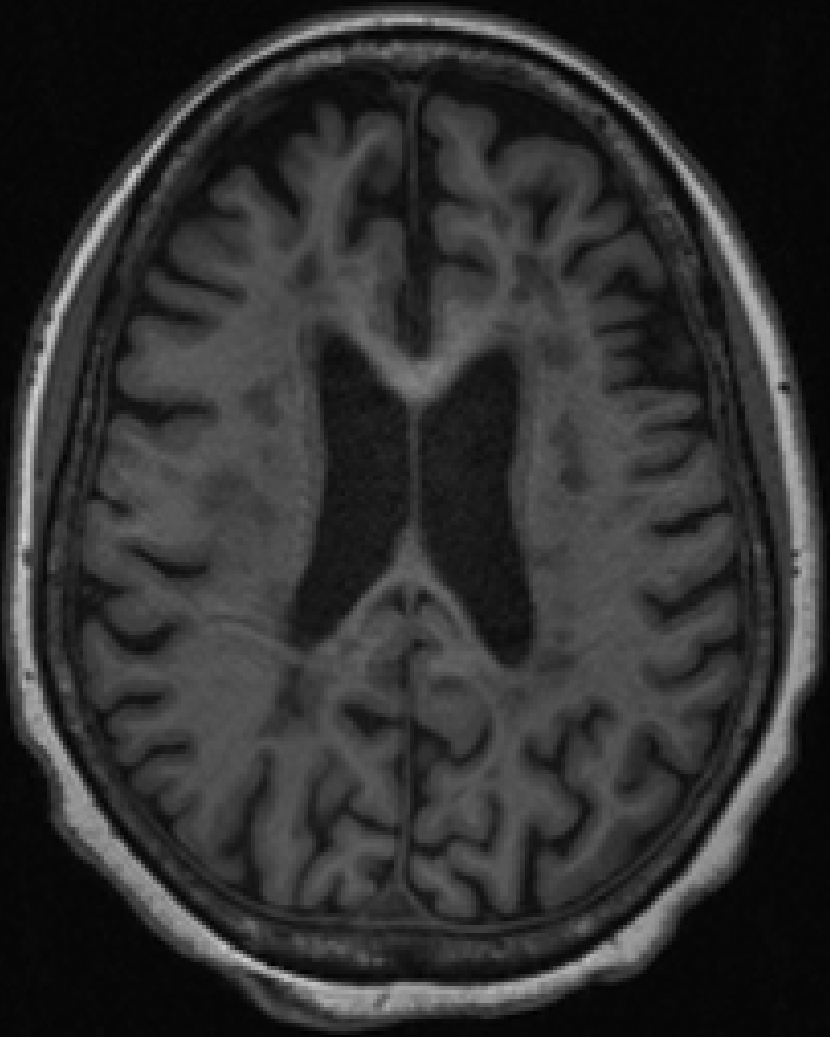} \label{fig:example_t1}}
	\subfloat{\includegraphics[width=.22\columnwidth]{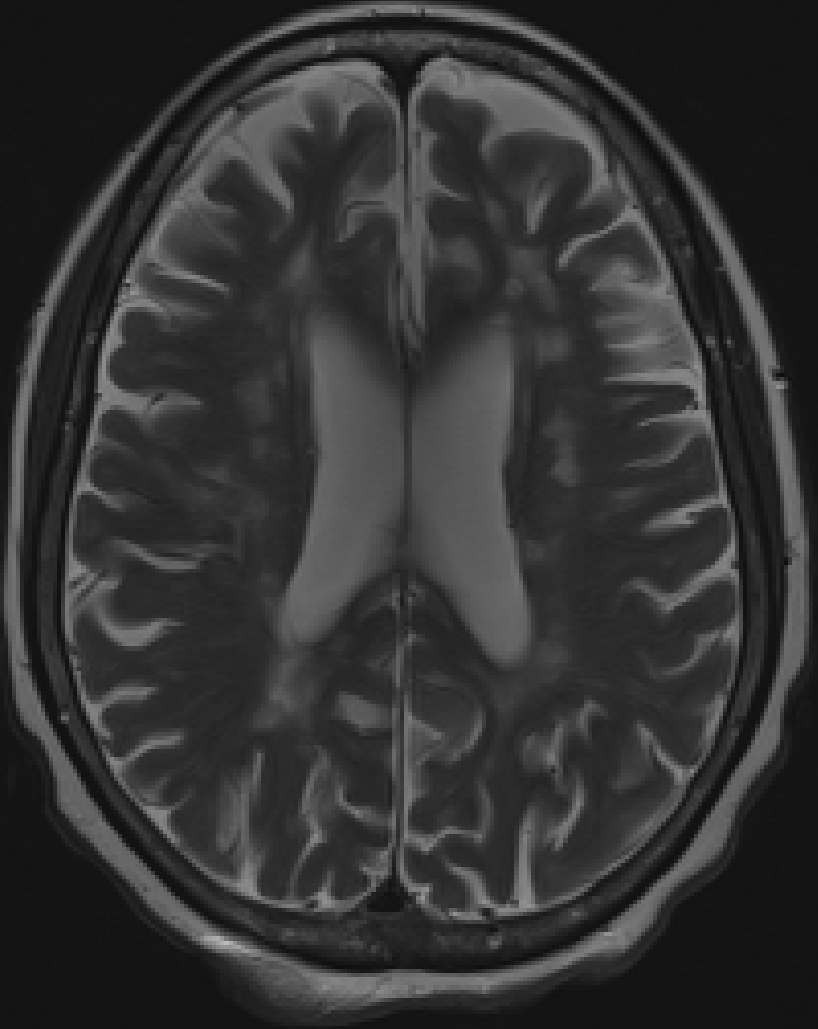} \label{fig:example_t2}}
	\subfloat{\includegraphics[width=.22\columnwidth]{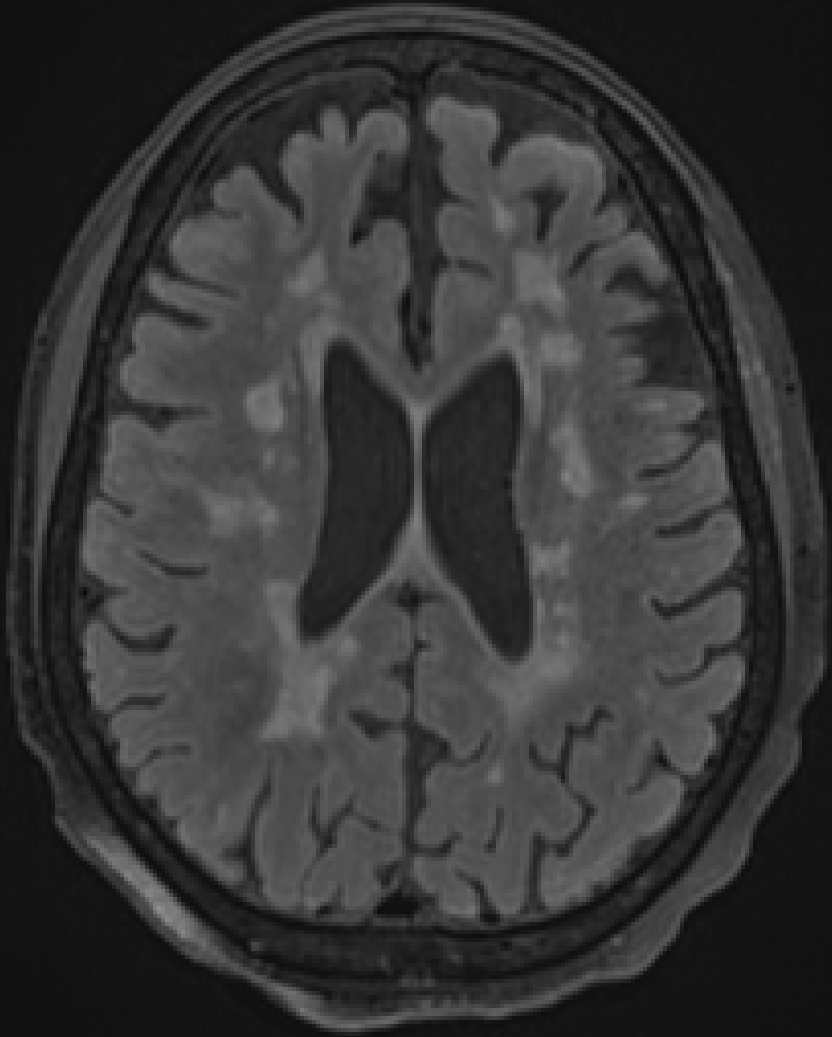} \label{fig:example_t2flair}}
	\subfloat{\includegraphics[width=.22\columnwidth]{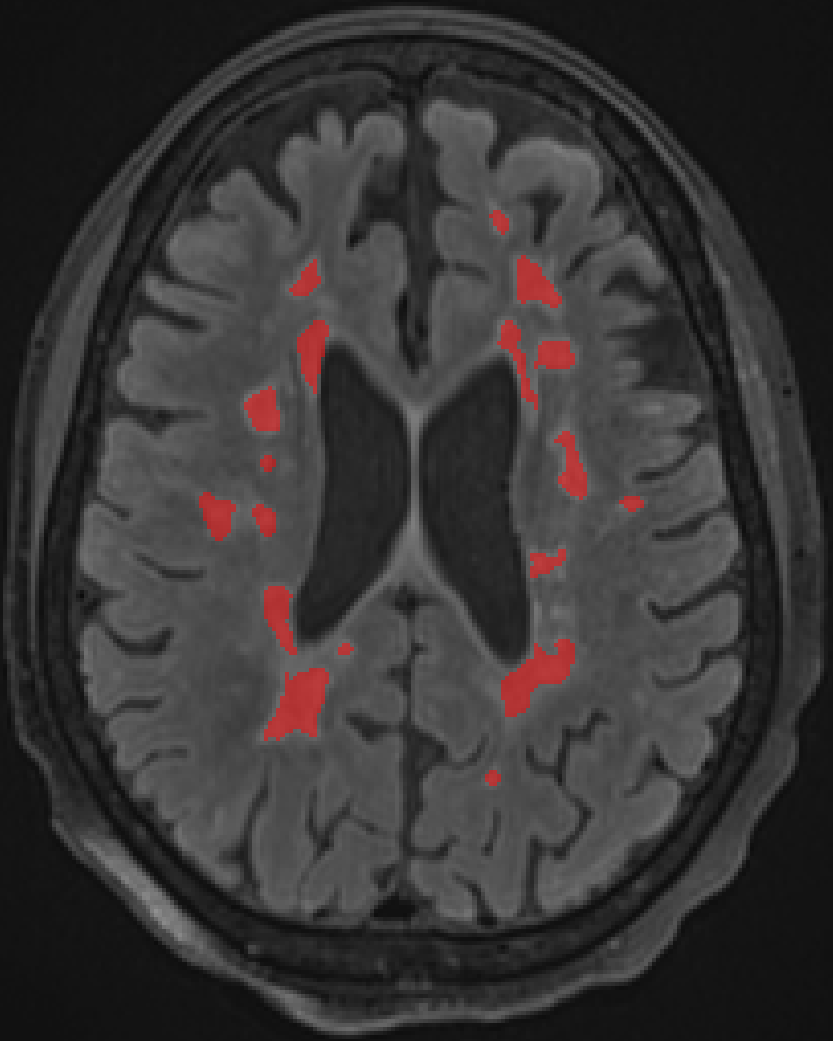} \label{fig:example_roi}}
	\caption{ 
	    Example T1, T2, T2-FLAIR images and corresponding mask traced by a human expert and marked in red. 
	}
	\label{fig:mri_examples}

\end{figure}

Multiple Sclerosis (MS) is an inflammatory demyelinating disease of the brain and spine characterized by the presence of hyperintense lesions on T2 weighted Magnetic resonance imaging (MRI) images  \cite{mslesionreview3}. 
Lesion volume measured on MRI is a clinically important marker of disease progression, which can be used to monitor and guide treatment. Currently, MS lesion segmentation is often performed manually or requires extensive manually editing, which is tedious, time-consuming and highly operator-dependent.
In the brain, MS lesions can vary greatly in terms of location, size, shape, and conspicuity on T2 weighted images (Fig.~\ref{fig:mri_examples})
Therefore, a fast and fully automated lesion segmentation tool is highly desired to improve accuracy and reproducibility while saving time in the clinic \cite{mslesionreview2}. 

In recent MS lesion segmentation algorithms, deep neural networks \cite{deepmedic,dlmslesionreview3,biolstm} which learn the mapping between images and lesion masks have achieved the best performance.
Patch based methods \cite{deepmedic,dlmslesionreview3} which extract small patches for classification of the central voxel suffers from slow training and ignores global brain structure information. 
Two-dimensional methods \cite{deepmedic,unet} that are based on in-plane segmentation does not utilize contextual information along the slice direction.
Three-dimensional convolutional neural networks (CNN) \cite{unet3d,biolstm} have been proposed to capture image dependencies between consecutive slices.
However, by trading-off between the convolution kernel size and the number of pooling layers, these methods can only capture limited local receptive field and short-range dependencies. 

kU-Net \cite{biornn} which combines U-Net \cite{unet} and recurrent neural network (RNN), is an effective approach to capture dependencies between different slices along the axial direction.
kU-Net can be further improved \cite{biolstm} by exploiting multi-modal MRI images and applying long short term memory (LSTM) network. 
However, these methods only capture the axial connections and ignore the sagittal and coronal directions in 3D images. 
Additionally, RNN and LSTM are inherently limited to short-range dependencies \cite{nmtatt1,nmtatt2,attisallyouneed}.  
Non-local neural networks \cite{nonlocal} incorporates a powerful self-attention mechanism \cite{attisallyouneed} to aggregate contextual information for every single pixel from all other pixels. 
The rich dependency information captured by this method has been shown to benefit various applications \cite{nonlocalderain,nonlocalaction} in image and video processing.
However, this non-local method needs to compute very large attention maps (e.g. 3D MRI image volume with size $512\times512\times256$ requires attention map with size $67108864\times67108864$ ), which greatly increases computational and memory demands, and also makes the model prone to overfitting in the case of limited samples.
Currently, leveraging the contextual and domain-specific MRI information for automatic MS lesion segmentation remains a non-trivial task.

\begin{figure}[!t]
\centering
\includegraphics[width=1.0\textwidth]{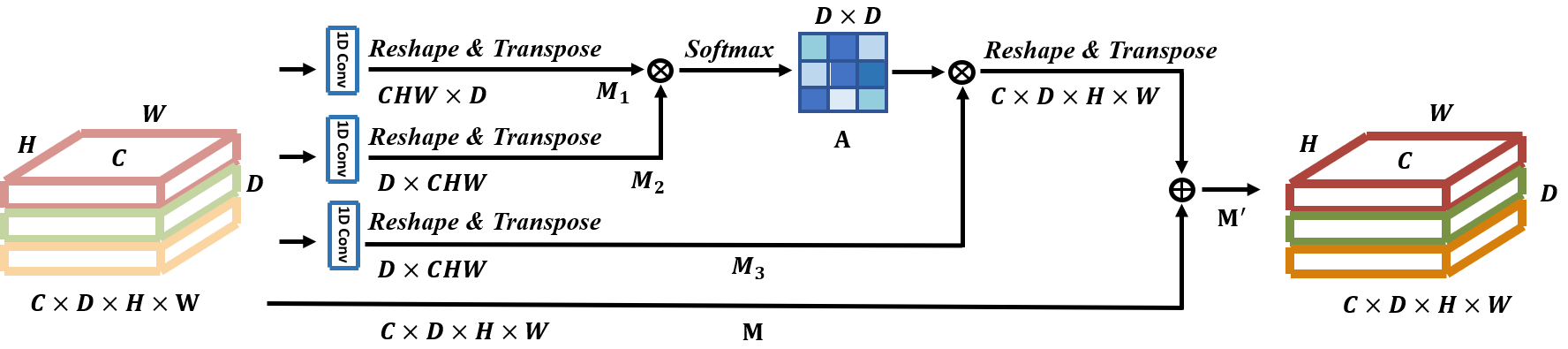}
\caption{Details of the SA block. 
} \label{fig:sa_block}

\end{figure}

The contributions of our RSANet are three folds.
First, unlike methods \cite{biornn,biolstm} using RNN or LSTM to capture the slice-wise dependencies, where RNN and LSTM have inherent drawbacks \cite{nmtatt1,nmtatt2,attisallyouneed} of capturing long-range dependencies, we propose a novel slice-wise attention module, called SA Block (see Fig.~\ref{fig:sa_block}) to compute the response at a slice as a weighted sum of the features from all slices along the same direction. 
Second, we propose a recurrent slice-wise attention module, named RSA block, to capture long-range dependencies from all voxels. 
The input features passed into RSA block will be recurrently fed into SA blocks along sagittal, coronal and axial directions (see Fig.~\ref{fig:recurrent_concept}). 
All three SA blocks share the same parameters, and each SA block will aggregate information from previous SA blocks and finally captures the voxel-wise dense dependencies. 
Third, RSANet is memory and computationally friendly.
Compared with the non-local method \cite{nonlocal}, our method reduces GPU memory consumption by at least $28 \times$ and the number of floating point operations by at least $100 \times$ in computing the attention maps.
We then demonstrate the efficiency and effectiveness of our RSANet on MS lesion segmentation in a dataset with 43 patients.

\section{Methodology}

In this section, we discuss the details of our proposed RSANet for MS lesion segmentation.
We will present the slice-wise Attention (SA) block, a basic building block of RASNet.
We will then show how to construct the recurrent slice-wise attention (RSA) block with SA block to capture the voxel-wise long-range dependencies.

\subsection{Slice-wise Attention Block}

Global brain structure and dependencies between lesion locations and brain areas are important features in MS, giving rise to sequential slice dependencies on MRI \cite{biornn,biolstm}.
In order to capture the long-range dependencies between different slices, we propose a slice-wise attention block, which is self-manipulated and could be put into any layer of an existing CNN architecture.

As shown in Fig.~\ref{fig:sa_block}, using SA block along the axial direction as an example, given a local feature map $ \mathbf{M} \in \mathbb{R}^{C\times D\times H\times W}$, we first apply a set of tensor transformations to the feature map $\mathbf{M}$.
Specifically, we reshape-transpose $\mathbf{M}$ to $ \mathbf{M}_1 \in \mathbb{R}^{CHW\times D}$, and reshape-transpose $\mathbf{M}$ to $ \{\mathbf{M}_2, \mathbf{M}_3\} \in \mathbb{R}^{D\times CHW}$ ($CHW$ means the product of $C, H, W$).
We then apply a matrix multiplication between $\mathbf{M}_2$ and $\mathbf{M}_1$ and followed by a softmax operation on the multiplication result to get the slice-wise attention map $\mathbf{A}$ as follows:
\begin{equation}
    \mathbf{A}_{ij} = \frac{\exp(\mathbf{M}_2[i,:] \mathbf{M}_1[:,j])}{\sum_{k=1}^{D}{\exp(\mathbf{M}_2[i,:] \mathbf{M}_1[:,k])}}~~,
    \label{eqn:attention_map}
\end{equation}
where $\mathbf{A}_{ij}$ measures the impact of $j^{th}$ slice on the $i^{th}$ slice, and each row of the attention map $\mathbf{A}$ is the weight that will be used to aggregate impact from all other slices. 
Also, $\mathbf{M}_2[i,:]$ denotes the $i^{th}$ row of $\mathbf{M}_2$, and $\mathbf{M}_1[:,j]$ denotes the $j^{th}$ column of $\mathbf{M}_1$.
We then attend $\mathbf{M}_3$ to $\mathbf{A}$ by another matrix multiplication, followed by a reshape-transpose (denoted as $RT()$ in equations) operation and an element-wise sum operation with original $\mathbf{M}$ to get the final output $\mathbf{M}^{'}$:
\vspace{-1ex}
\begin{equation}
    \mathbf{M}^{'}= \alpha RT(\mathbf{A}\mathbf{M}_3) + \mathbf{M},
    \label{eqn:attended_sum}
\end{equation}
where $\alpha$ is a scaling parameter that will be updated with other parameters through back-propagation.
Through Eqn.~\eqref{eqn:attention_map} and Eqn.~\eqref{eqn:attended_sum}, we can see that the output of SA block is a linear combination of input feature map and features aggregated from the weighted sum of all other slices, which captures the long-range dependencies between slices.
Unlike previous work \cite{biornn,biolstm} which uses RNN and LSTM with inherent short-range problems \cite{nmtatt1,nmtatt2}, we model slice dependencies by exploring spatial contextual information through all slices.
Besides, with the recurrent module introduced in the next section, we can fuse the information efficiently from three slice directions.

\begin{figure}[!t]
\centering
\includegraphics[width=0.75\textwidth]{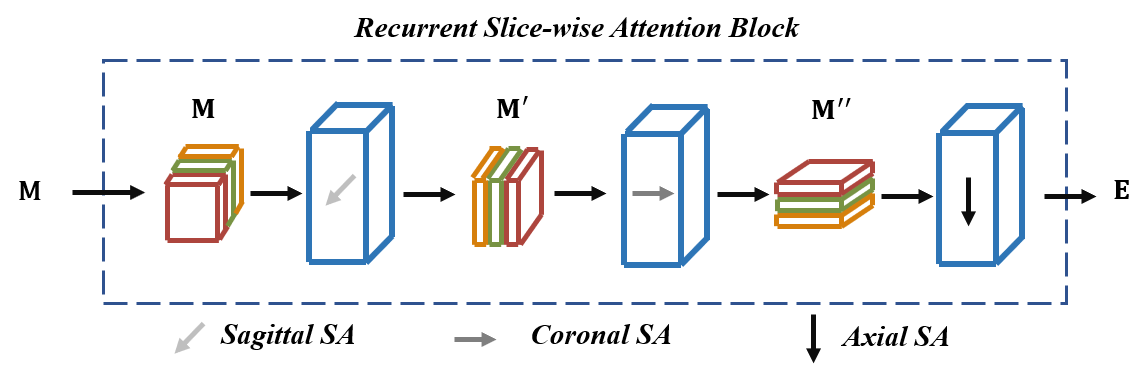}
\caption{Details of the RSA block. 
    The input is the feature map $\mathbf{M}$. 
    RSA block takes $\mathbf{M}$ and recurrently produces $\mathbf{M}^{'}$, $\mathbf{M}^{''}$ as intermediate results.
    Finally, RSA block will output $\mathbf{E}$, where each voxel is a weighted sum of all other voxels.
} \label{fig:rsa_block}
\end{figure}

\subsection{Recurrent Slice-wise Attention Block}

Dense contextual dependency information helps regularize the gradient information propagation through the whole network, and in our case, this regularization could help the deep network understand the complicated distributions of MS lesions as well as the relationship between brain structure and lesion masks.
Therefore, it is imperative to capture the long-range dependencies among all voxels and in the meanwhile avoid intolerable memory and computation consumption brought by non-local block \cite{nonlocal}. 
We propose our recurrent slice-wise attention (RSA) block, which combines the long-range dependencies among three slice directions in a recurrent manner.

The overall RSA block is shown in Fig.~\ref{fig:rsa_block}.
The RSA block consists of three SA blocks with different slice directions i.e. sagittal, coronal and axial.
Each SA block in the RSA block share the same convolutional kernel parameters, but their weight parameter $\alpha$ is updated independently.
In the first recurrent loop, the RSA block takes the input feature $\mathbf{M}$ extracted by previous convolutional layers and outputs $\mathbf{M}^{'}$ where the image slices are sagittally attended. 
We then repeat the other two SA blocks on coronal and axial directions and finally get the output $\mathbf{E}$ which is attended densely among all voxels. 
The $\mathbf{M}, \mathbf{M}^{'}, \mathbf{M}^{''}, \mathbf{E}$ have the same shape, and thus this RSA block can be put anywhere in the convolutional network.

\begin{figure}[!t]
\centering
\includegraphics[width=0.95\textwidth]{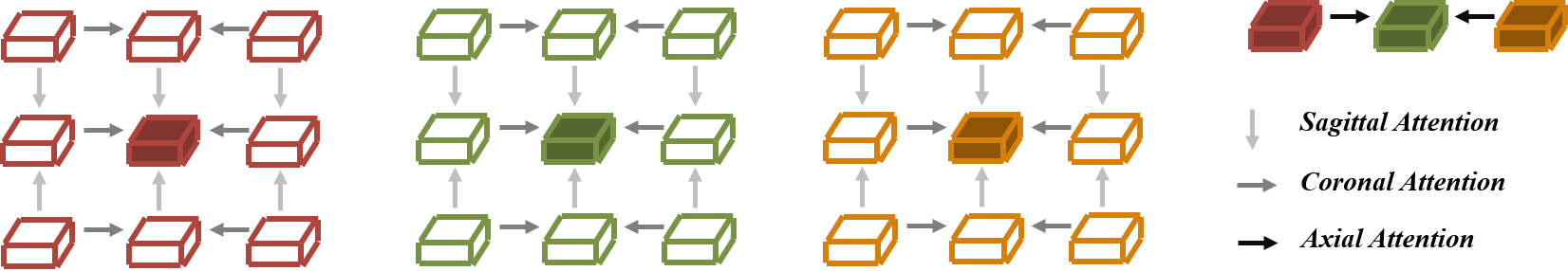}
\caption{An example of information propagation in RSA block.
} \label{fig:recurrent_concept}
\end{figure}

Single SA block can only capture the slice-wise dependencies along one direction, so we recurrently perform SA block on three slice directions to make up the deficiency of single SA block and obtain the global voxel-wise contextual information.
Details of the information propagation for our RSA block is shown in Fig.~\ref{fig:recurrent_concept}.
Suppose the input of the RSA block is $\mathbf{M} \in \mathbb{R}^{3\times 3\times 3}$, and the red cubes denote the first slice of $\mathbf{M}$ along the axial direction (green for the second slice and yellow for the third slice). 
We show how adjacent voxels aggregate the central voxel in the RSA block and similar procedures can be applied to all the other voxels.
In the first SA block, $\mathbf{M}^{'}[1,1,1] \leftarrow SA1(\mathbf{M}[1,0,1],\mathbf{M}[1,2,1]) + \alpha_0 \mathbf{M}[1,1,1])$, where $SA1()$ denotes the weighted sum of $\mathbf{M}[1,0,1],\mathbf{M}[1,2,1]$ and the weighting matrix is obtained by the attention map of this SA block.
Then in the second SA block, $\mathbf{M}^{''}[1,1,1] \leftarrow SA2(\mathbf{M}^{'}[1,1,0],\mathbf{M}^{'}[1,1,2]) + \alpha_1 \mathbf{M}^{'}[1,1,1])$, where $SA2()$ is the same function as $SA1()$ but acts on different inputs and weights.
Finally, we can obtain $\mathbf{E}[1,1,1] \leftarrow SA3(\mathbf{M}^{''}[0,1,1],\mathbf{M}^{''}[2,1,1]) + \alpha_2 \mathbf{M}^{''}[1,1,1])$.

In general our RSA block help gather global contextual information by weighted sum from all voxels,  and in the meanwhile dramatically reduce the computational cost and memory usage when computing the attention map.

\section{Experimental Results}

We demonstrate the performance of the proposed RSANet on MRI images acquired with three different contrast (T1, T2, and T2FLAIR) on a GE 3T scanner.
A total of 43 MS patients were included in this study.
The size of each MRI image dataset varies from $230 \times 320 \times 44$ to $260 \times 320 \times 60$ with a voxel size of $0.7\times 0.7 \times 3.0~mm$.
Images were linearly co-registered using FSL neuroimaging toolbox (FLIRT command).
MS lesions were manually segmented by a neural radiologist with six-year experience.

\subsection{Training, Testing and Implementation Details}

The proposed RSANet was implemented using PyTorch and 3D U-Net \cite{unet,unet3d} is used as our backbone network structure.

\subsubsection{Loss Function.}
The data imbalance problem is critical in MS lesion segmentation, as the ROIs usually compose a tiny portion of the MRI images. 
In our dataset, the rate of MS lesion is only $5.15 \cdot 10^{-4}$.
Therefore we adopted an exponentially weighted CrossEntropy as our loss function.
The weight for MS lesion ROIs is $\dfrac{e^r}{e^r + e^{1-r}}$ and the weight for others is $\dfrac{e^{1-r}}{e^r + e^{1-r}}$, where $r$ is on average the portion of MS lesion in the MRI images.

\subsubsection{Training parameters.}
We randomly split our dataset into two subset, 20 images for training and the rest 23 for testing. 
We evaluated each method on five random splits and take the average as the result to compare performance.
As the image size varies from patient to patient, we performed random crop with fixed cropping size on the original images to make sure they were in the same size for training.
ADAM \cite{adam} with the initial learning rate of $1e^{-3}$ and weight decay of $1e^{-5}$ was used to train each method.
Each method was trained with a batch size of $4$ on a single Titan XP GPU, and training would stop after 800 iterations.

\subsubsection{Implementation Details.}
As both non-local block and RSA block could be put anywhere in the network, we tried three different ways of putting these blocks and compare their performance. 
Non-local block can only be put in the layers close to the bottom of 3D U-Net \cite{unet,unet3d}, as it would cause serious memory problem if it is put in the higher layers.
Therefore, for fair comparison, we choose three different ways to put these blocks: 1) Single bottom layer, denoted as NCL-010 and RSA-010; 2) Encoder and decoder of the second bottom layer, denoted as NCL-101 and RSA-101; 3) All three places mentioned above, denoted as NCL-111 and RSA-111.

\subsection{Quantitative and Qualitative Results}

We compared our models with several advanced approaches, including 3D U-Net \cite{unet,unet3d} and different settings of non-local neural networks \cite{nonlocal}. 
For fair comparison, we obtained these methods from public implementations and adjusted their parameters to get the best performance.
We used Dice similarity coefficient (Dice) and Intersection of Union (IoU) as our evaluation metrics.
As the number MS lesion vary greatly from patient to patient, simple average of all samples on Dice and IoU would cause bias on evaluating the real performance. 
Accordingly, we proposed a new evaluation method, in which all voxels from all samples were pooled to compute ``voxel average'' Dice and IoU as opposed to the conventional sample average.

\subsubsection{Quantitative Results.}
As shown in Table.~\ref{table:results}, when 3D U-Net equipped with non-local block, it outperformed original 3D U-Net in all metrics regardless of where we put the non-local block.
Especially in the case of NCL-010, it outperformed 3D U-Net by $0.83 \%$ of Voxel Avg. Dice and $1.0 \%$ of Voxel Avg. IoU, the result of which is consistent as reported in non-local method \cite{nonlocal}.
RSANet outperformed both 3D U-Net and non-local net with different non-local positions in all metrics (Table.~\ref{table:results}).
Furthermore, the sample average Dice and IoU score of RSANet increased with the number of RSA blocks in the network. 
This property is beneficial as our RSA blocks barely increase the cost of floating computations and memory usage.

\begin{table}[!t]
\centering
\caption{Quantitative comparison of MS lesion segmentation}
\resizebox{1.0\columnwidth}{!}
{
    \begin{tabular}{|c|c|c|c|c|c|}
        \hline \hline
        \tabincell{c}{Method}            &
        \tabincell{c}{Sample Avg. Dice}       &
        \tabincell{c}{Voxel Avg. Dice}       &
        \tabincell{c}{Sample Avg. IoU}       &
        \tabincell{c}{Voxel Avg. IoU}      \\
        \hline
        3D U-Net \cite{unet,unet3d}              & 63.984\%      & 69.640\%      & 48.754\%      & 53.506\%      \\
        \hline
        NCL-010 \cite{nonlocal}                  & 64.346\%      & 70.473\%      & 49.231\%      & 54.500\%      \\
        NCL-101 \cite{nonlocal}                  & 64.069\%      & 70.121\%      & 49.103\%      & 54.090\%      \\
        NCL-111 \cite{nonlocal}                  & 64.074\%      & 70.185\%      & 48.833\%      & 54.170\%      \\
        \hline
        RSA-010               & 65.300\%      & 70.207\%      & 50.248\%      & 54.200\%      \\
        RSA-101               & 65.949\%      & \bf{71.589}\% & 50.896\%      & \bf{55.847}\% \\
        RSA-111               & \bf{66.011}\% & 71.054\%      & \bf{50.917}\% & 55.201\%      \\
        \hline \hline
    \end{tabular} 
}
\label{table:results}

\end{table}

\subsubsection{Qualitative Results.}
\begin{figure}[!t]
	\centering
	\subfloat{\includegraphics[width=.22\columnwidth]{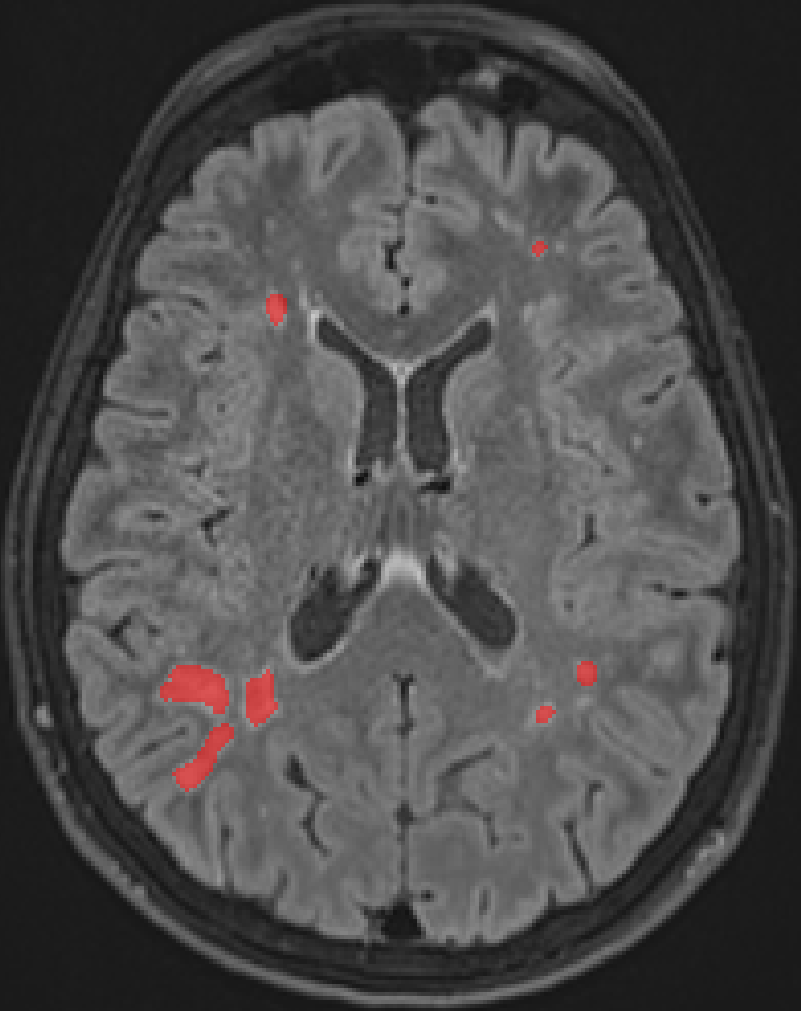} \label{fig:1_roi}}
	\subfloat{\includegraphics[width=.22\columnwidth]{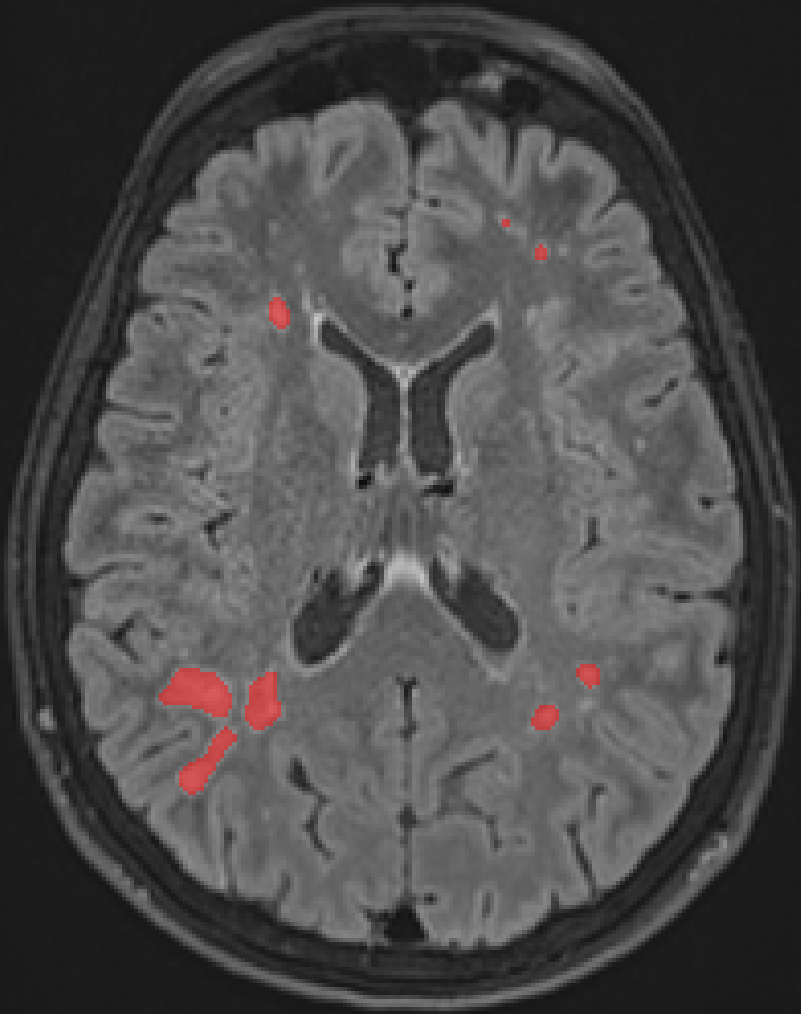} \label{fig:2_rsa111}}
	\subfloat{\includegraphics[width=.22\columnwidth]{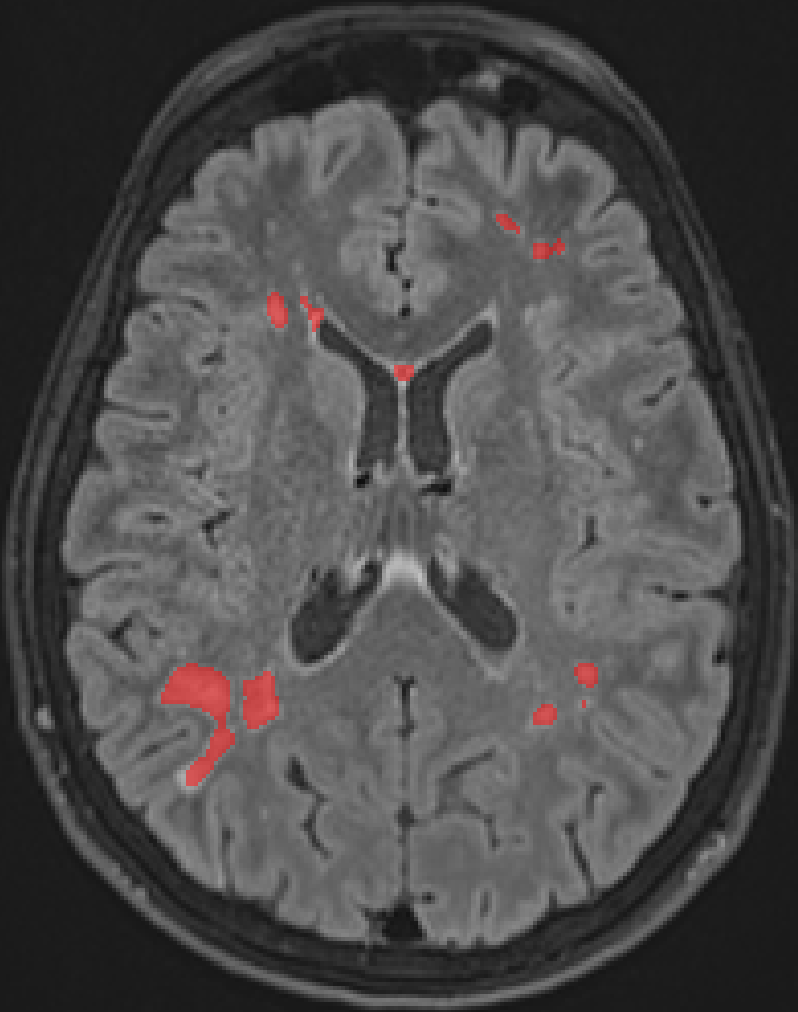} \label{fig:3_ncl010}}
	\subfloat{\includegraphics[width=.22\columnwidth]{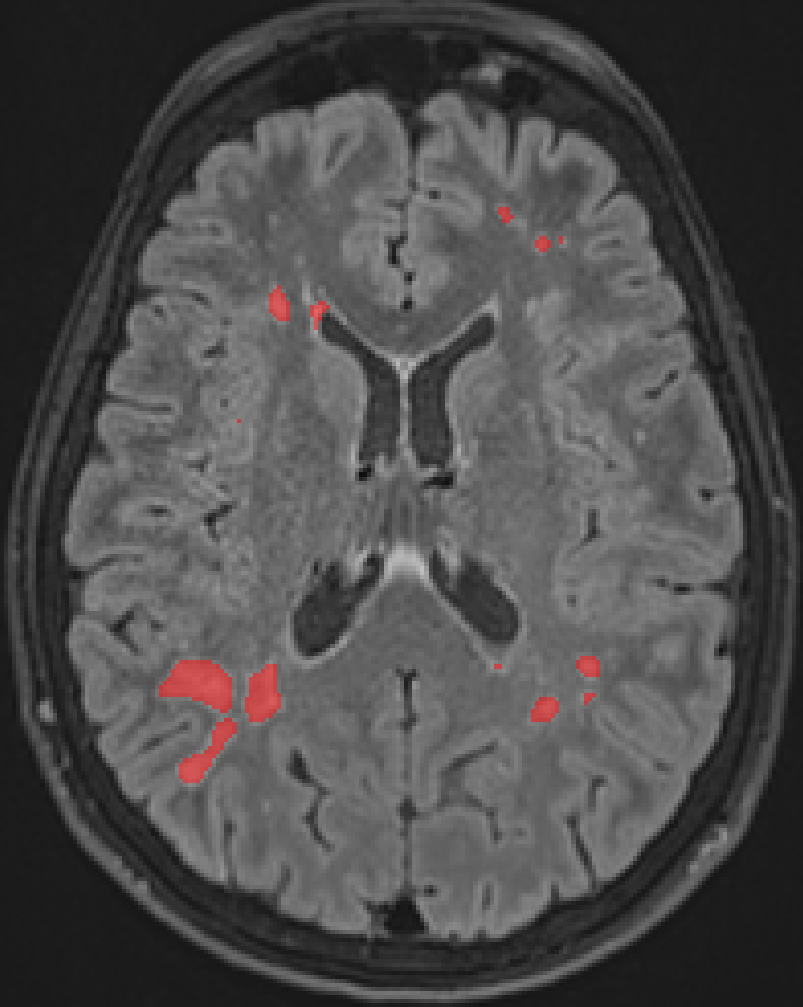} \label{fig:4_ncl000}}
	\caption{ 
	    Example segmentation result.
	    From left to right are ground truth label, results of RSA-111, NCL-010 and 3D U-Net. 
	}
	\label{fig:qualitative_result}

\end{figure}

We choose one slice from a testing image, and compare the qualitative results of different models with ground truth labels.
As we can see from Fig.~\ref{fig:qualitative_result}, since both 3D U-Net and non-local network are not able to efficiently capture the long-range dependencies between MS lesions and brain structure, they suffer from an over-segmenting problem. 

\vspace{-1ex}
\section{Conclusions}
\vspace{-1ex}
We presented a novel recurrent slice-wise attention network, which incorporates three slice-wise attention blocks recurrently.
Our proposed method can capture the long-range dependencies within the MRI images for MS lesion patients, which exploit the contextual information between the brain structure and lesion masks.
Our method not only achieves the high accuracy on MS lesion segmentation tasks, but also reduces dramatically the computational cost and GPU memory usage.
Experimental results showed that our method outperformed other state-of-the-art methods.
Our method can be put anywhere in the deep network and thus has the potential for other 3D medical image segmentation tasks.



\bibliographystyle{splncs04}
\bibliography{mybibliography}

\end{document}